\documentclass[%
 reprint,
superscriptaddress,
 amsmath,amssymb,
 aps,
pra,
longbibliography
]{revtex4-2}

\usepackage{graphicx}% Include figure files
\graphicspath{{./pdf_eps/}}
\DeclareGraphicsExtensions{.eps,.png}

\usepackage{dcolumn}% Align table columns on decimal point
\usepackage{bm}% bold math
\usepackage{natbib}
\usepackage{textcase}
\usepackage{url}% add hypertext capabilities
\usepackage{amsmath}
\usepackage{bbold}
\usepackage{calc}%     needed for the width/height calculations
\usepackage{dirtytalk}% needed for quotation marks
\usepackage{mathtools}
\usepackage{array}
\usepackage{siunitx}
\usepackage{hhline}
\usepackage{xcolor}
\usepackage{soul}
\usepackage{float}
\setstcolor{red}

\begin{document}

\preprint{APS/123-QED}

\title{First-passage time to capture for diffusion in a 3D harmonic potential}% Force line breaks with \\
%\thanks{A footnote to the article title}

\author{Tianyu Yuan}
 %\altaffiliation[Also at ]{Physics Department, Yale University.}%Lines break automatically or can be forced with \\
\affiliation{%
    Integrated Graduate Program in Physical and Engineering Biology, Yale University, New Haven, Connecticut 06520, USA
}
\affiliation{%
    Department of Physics, Yale University, New Haven, Connecticut 06520, USA
}%

\author{Ivan Surovtsev}
\affiliation{%
    Department of Physics, Yale University, New Haven, Connecticut 06520, USA
}
\affiliation{
    Department of Cell Biology, Yale School of Medicine, New Haven, Connecticut 06520, USA
}

\author{Megan C. King}
\affiliation{%
    Integrated Graduate Program in Physical and Engineering Biology, Yale University, New Haven, Connecticut 06520, USA
}
\affiliation{
    Department of Cell Biology, Yale School of Medicine, New Haven, Connecticut 06520, USA
}
\affiliation{
    Department of Molecular, Cell and Developmental Biology, Yale University, New Haven, Connecticut 06511, USA
}

\author{Simon G. J. Mochrie}
\email{simon.mochrie@yale.edu}
\affiliation{%
    Integrated Graduate Program in Physical and Engineering Biology, Yale University, New Haven, Connecticut 06520, USA
}
\affiliation{%
    Department of Physics, Yale University, New Haven, Connecticut 06520, USA
}
\affiliation{%
    Department of Applied Physics, Yale University, New Haven, Connecticut 06520, USA
}

\date{\today}

\begin{abstract}
% Reaction dynamics of diffusion-controlled processes can often be characterized by their first passage time. 
% The first passage time problem of a harmonically trapped particle generalizes to many important questions, of most interest is the polymer loop formation, as its relevance in protein and RNA folding, and DNA loop formation.

 We determine the survival probability and
first-passage time (FPT) to capture
for a harmonically trapped particle, diffusing outside an absorbing spherical boundary
by directly solving the differential equation for the survival
probability.
This solution,
obtained as an infinite sum over the relevant eigenfunctions,
corrects previously published results
[D. S. Grebenkov, J. Phys. A 48, 013001 (2014)].
To verify our calculations, we perform simulations of the survival probability,
that accurately reproduce the analytic solutions for a range of parameter values.
We then obtain the corresponding FPT distribution as the negative time derivative of the survival
probability.
Finally, we derive an expression for mean first-passage time (MFPT), also
as a sum over eigenfunctions.
Numerical evaluation of the first twenty-five terms in this sum closely
matches the MFPT obtained by a different method
in D. S. Grebenkov, J. Phys. A 48, 013001 (2014).
We also find that,
in the limit of vanishing trap stiffness,
the amplitude of the first term in our infinite-sum solution for the survival probability matches the theoretical escape probability for the potential-free diffusion-to-capture process.

% We verify our theoretical results b

% 0. first passage time problem of relevant in physical, chemical and biological processes.
% 1. of particular interest are the systems governed by harmonic potentials, which are relevant for polymer loop formation.
% 2. revisit the solution of survival probability and mean first passage time provided by Grebenkov, and produce the correct solution in the series form.
% 3. 
% 4. 

\end{abstract}

\maketitle

\section{Introduction}
\label{intro}
Diffusion-to-capture of a particle subject to a
harmonic potential is often one of the simplest theoretical representations of
myriad processes, and has been invoked to describe
protein folding \cite{van1992,manhart2014,hartich2019},
polymer cyclization \cite{szabo1980,toan2008,jeon2014},
neuronal firing activity \cite{ricciardi1979,van1992,lanska1994},animal searching behavior~\cite{ebeling1999,campos2014,costa2024}, and financial phenomena~\cite{chicheportiche2014}.

As a result, there has been longstanding theoretical interest
in calculating the first-passage time (FPT) to capture in this scenario
~\cite{pontryagin1933,szabo1980,redner2001,toan2008,metzler2014,li2019}.
Recently,
Grebenkov~\cite{grebenkov2014} derived the survival probability and the corresponding FPT distribution for
a particle in a harmonic potential, diffusing either inside (interior problem) or outside (exterior problem) an absorbing boundary for arbitrary
spatial dimensions.
Unfortunately, the solution to the 3D exterior
problem given in Ref.~\cite{grebenkov2014} is incorrect.
In this paper, we provide the correct solution,
revealing the following:
(1) For particles with initial, near-equilibrium separations
from the origin, the FPT
distribution is nearly single-exponential.
(2) For particles with initial separations,
significantly larger than the equilibrium separation,
there is an ``incubation time'', during which the particle
moves from its initial distant separation to the
equilibrium separation.
Once the particle is at the equilibrium separation,
it then undergoes the near single-exponential first-passage
process.
(3) For particles with initial separations,
significantly smaller than the equilibrium
separation,
the particles is either rapidly captured or it
enjoys a temporary reprieve by escaping to the
equilibrium separation, whereupon
it then undergoes the near single-exponential first-passage
process.

Following Ref.~\cite{grebenkov2014},
we solve the differential equation obeyed by the survival
probability, subject to an absorbing boundary condition
at the capture radius and the initial condition that the particle is
at a given radius.
Thus,
we obtain a corrected expression for the survival probability as a function of the initial separation and time, as an infinite series of confluent hypergeometric (Tricomi) functions. 
Each Tricomi function is multiplied by
an exponentially-decaying-in-time factor with 
a time constant proportional to a zero of the Tricomi function. 
The corresponding FPT distribution follows immediately
by differentiation of the survival probability.

By comparing our analytical results with simulations of an overdamped bead in a harmonic potential, 
we demonstrate excellent agreement between the theoretical and simulated survival probabilities across a range of initial separations and potential trap stiffnesses, thereby validating our analytic series solution.

We also obtain the mean first-passage time (MFPT) as a function of the
particle's initial radial coordinate. This result shows excellent
agreement with the MFPT from Ref.~\cite{grebenkov2014}, obtained by
a different method as an integral.

% to the simulation result of an overdamped bead connected to a fixed spring, we show agreement between our analytical solutions to the survival probability and the simulated survival probability for multiple values of initial separations and potential strength.

% When the time scale of a Gaussian polymer's end-to-end loop formation is much larger than the internal chain relaxation time, 

% SSS and Thirimulai have shown that it can be approximated by Smoluchowski equation using single time scale.

\section{Theory}
\label{theory}
%
%\subsection{Particle diffusing in a potential}
%\label{diffusion}
%
Ignoring inertia, 
the positional distribution function, $p({\bf x},t)$,
of a particle, 
%the particle's position distribution function,
diffusing  in a spherically-symmetric, harmonic potential
of stiffness, $k$, $V({\bf x})=\frac{1}{2} k r^2$, obeys the Smoluchowski equation
($r$ is the radial coordinate): %(forward Fokker-Planck equation):
\begin{equation}
    \frac{\partial p}{\partial t}
    =D\nabla^2p +\frac{1}{\zeta}\nabla\cdot(p\nabla V)
    =
    D\nabla^2p +\frac{k}{\zeta}\nabla\cdot(p{\bf x}),
    \label{forwFP}
\end{equation}
where $D$ and $\zeta$ are the particle's diffusivity and friction coefficient, respectively, which are related via
the Einstein relation, $D=\frac{k_BT}{\zeta}$.
Since we are interested in the FPT for the particle to reach a certain radial coordinate, $r=L$,
we impose an absorbing boundary condition
at $r=L$
\begin{equation}
    p(L,\theta, \phi, t)=0,
    \label{EQ2}
\end{equation}
which ensures that $p$ corresponds to trajectories that have not
reached $r=L$.
Solving Eq.~\ref{forwFP} subject to this boundary condition and the initial condition that the particle
is at ${\bf x}_0$ at $t=0$, {\em i.e.}
\begin{equation}
  p(\mathbf{x},0)=\delta({\bf x}-{\bf x}_0), 
\end{equation}
%is often called the ``forward problem'', and
yields the conditional
probability that the particle is at ${\bf x}$ at time, $t$, given that it was at ${\bf x}_0$ at time zero, namely
$p({\bf x}, t|{\bf x}_0,0)$. The conditional probability
that the particle survives until time, $t$, irrespective of
its position at time $t$ given that it starts at ${\bf x}_0$
-- the ``survival probability'' -- is obtained from
$p({\bf x}, t| {\bf x}_0,t_0)$ follows
from marginalizing ${\bf x}$:
\begin{equation}
    S(t|r_0) = \int_L^\infty dr ~ r^2  \int d \Omega ~ p({\bf x}, t| {\bf x}_0,0)
    \label{EQ4}
\end{equation}

Because of the angular integral in Eq.~\ref{EQ4},
the survival probability depends only on the
particle's  initial radius and not on its initial
angles.
In fact, as was shown previously in Ref.~\cite{szabo1980},
the survival
probability obeys the following two-variable,
partial differential equation
\begin{equation}
    \frac{\partial S}{\partial t}=D\frac{\partial^2 S}{\partial r_0^2}+(\frac{2D}{r_0}-\frac{kr_0}{\zeta})\frac{\partial S}{\partial r_0},
    \label{mainEQ}
\end{equation}
subject to the boundary condition,
\begin{equation}
    S(t|L)=0,
    \label{bcS}
\end{equation}
and the initial condition
\begin{equation}
    S(0|r_0)=1.
    \label{icS}
\end{equation}
% 
%For brevity,
%henceforth we will write $r$
%instead of $r_0$.

To solve Eq.~\ref{mainEQ},
we assume a separable solution:
$S(t|r_0)=R(r_0)T(t)$, which we substitute
into Eq.~\ref{mainEQ},
resulting in the
following equations for $T$
and $R$:
\begin{equation}
    \frac{dT}{dt}+\lambda T=0, 
    \label{timeEQ}
\end{equation}
and
\begin{equation}
    D\frac{d^2R}{dr_0^2}+(\frac{2D}{r_0}-\frac{kr_0}{\zeta})\frac{dR}{dr_0}+\lambda R=0,
    \label{spatialEQ}
\end{equation}
where $\lambda$ is the
separation-of-variables constant.

Clearly, the solution for $T$ is
\begin{equation}
    T=e^{-\lambda t}.
\end{equation}
In order that $T$, and therefore $S$, decay to zero as $t$ goes to infinity, as required on physical grounds, $\lambda$
must be positive.

The general solution to Eq.~\ref{spatialEQ}
is a confluent
hypergeometric function ~\cite{abramowitz1965}.
In fact, in order that $S$, and therefore $R$, remains finite at large $r_0$,
the required solution
is the
confluent
hypergeometric function of the second kind,
a.k.a. the Tricomi function, $U$~\cite{grebenkov2014}:
\begin{equation}
    R(r_0)=U\left(-\frac{\zeta\lambda}{2k},\frac{3}{2},\frac{kr_0^2}{2\zeta D}\right)
    =U\left(-\alpha,\frac{3}{2},\kappa z^2\right),
    \label{hyperU}
\end{equation}
where we introduced
the dimensionless
quantities
\begin{equation}
    \alpha=\frac{\zeta\lambda}{2k},
\end{equation}
\begin{equation}
    \kappa=\frac{kL^2}{2\zeta D},
\end{equation}
and
\begin{equation}
    z=\frac{r_0}{L}.
\end{equation}
To satisfy the boundary condition,
$R(L)=0 $ ({\em i.e.} Eq.~\ref{bcS}),
we require $\alpha=\alpha_n$, where $\alpha_n$ is the $n$-th zero of $U(-\alpha,3/2,\kappa)$.
It follows that
$\lambda$ can only take discrete values,
$\lambda_n=2\alpha_nk/\zeta$.

Equation~\ref{spatialEQ} can be rewritten
in Sturm-Liouville form,
\begin{equation}
    D\frac{d}{dr_0} \left ( r_0^2 e^{-\frac{k r_0^2}{2 \gamma D}} \frac{dR}{dr_0} \right )
    + \lambda r_0^2 e^{-\frac{k r_0^2}{2 \gamma D}}  R=0,
    \label{spatialEQ2}
\end{equation}
or
\begin{equation}
    \frac{d}{dz} \left ( z^2 e^{-\kappa z^2} \frac{dR}{dz} \right )
    + 4\alpha\kappa z^2 e^{-\kappa z^2}  R=0,
    \label{spatialEQ3}
\end{equation}
in terms of dimensionless quantities.
Sturm-Liouville theory, therefore, informs us that the functions,
\begin{equation}
\psi_n(z) = \frac{1}{N_n} U(-\alpha_n,\frac{3}{2},\kappa z^2),
\label{eigenfunctions}
\end{equation}
indexed by the positive integer, $n$,
constitute a complete set of orthogonal eigenfunctions,
%where $N_n$ is a normalization factor,
each with a unique
eigenvalue, $\lambda_n$ (Eq. \ref{spatialEQ2}) or $\alpha_n$ (Eq.~\ref{spatialEQ3}).
The $\psi_n$ are orthogonal with respect to
the weight function (in dimensionless form),
%\begin{equation}
 %   w(r_0)=r_0^2 e^{-\frac{k r_9^2}{2 \gamma D}},
%\end{equation}
%or
\begin{equation}
    w(z)=z^2 e^{-\kappa z^2},
\end{equation}
so that
\begin{equation}
    \int_1^\infty z^2 e^{-\kappa z^2}
    \psi_n(z) \psi_m(z)
  %  U(-\alpha_n,3/2,\kappa z)U(-\alpha_m,3/2,\kappa z)
    =\delta_{n,m},
\end{equation}
where $\delta_{m,n}$ is the Kronecker delta.
In Eq.~\ref{eigenfunctions}, $N_n$ is a normalization factor, given by
\begin{equation}
    N_n=\sqrt{\int_1^\infty z^2 e^{-\kappa z^2} [U(-\alpha_n, \frac{3}{2}, \kappa z^2)]^2 \, dz }.
    \label{normalization}
\end{equation}

Thus, the general solution for $S$, for the specified boundary conditions (Eq.~\ref{bcS}), is
given by
\begin{equation}
    S(t|z)=\sum_{n=1}^\infty c_n \psi_n(z)
    e^{-\lambda_n t}
     =\sum_{n=1}^\infty c_n \psi_n(z)
    e^{-\frac{2 k}{\zeta}  \alpha_n t},
    \label{fullsoln}
\end{equation}
where the $c_n$'s are arbitrary constants.
The quantity, $\frac{k}{\zeta}$, is the particle's relaxation rate in the harmonic potential.

To calculate the values of the $c_n$
for the specified initial condition
(Eq.~\ref{icS}),
we have
\begin{equation}
    S(0|z)=1=\sum_{n=1}^\infty c_n \psi_n(z).
    \label{ic1}
\end{equation}
% 
%for $z\in[1,\infty)$.
Multiplying by $z^2 e^{-\kappa z^2} \psi_m(z)$ on both sides of Eq.~\ref{ic1} and integrating over $z$ from 1 to $\infty$, we find:
\begin{equation}
    c_m = \int_1^\infty z^2 e^{-\kappa z^2}\psi_m(z)\,dz.
    \label{weight}
\end{equation}
Therefore, the solution for $S$, subject to the boundary and initial conditions given in Eqs.~\ref{bcS} and \ref{icS}, respectively, is given by Eq.~\ref{fullsoln},
where the eigenfunctions, $\psi_n(z)$, are given by Eq.~\ref{eigenfunctions}, with
$N_n$ given by Eq.~\ref{normalization}, and the eigenmode amplitudes,  $c_n$,
are given by Eq.~\ref{weight}.
These collected equations correct analogous expressions given in
Ref.~\cite{grebenkov2014}.
A Mathematica (Wolfram Research, Champaign, IL) notebook that calculates $\alpha_n$,
$N_n$, and $c_n$ for given $\kappa$ and $z$
is included in the Supplementary Information.

%\subsection{First passage time (FPT) distribution and mean first passage time (MFPT)}
%\label{FPT}

The first-passage time (FPT) distribution, $P(t|z)$, is obtained from  Eq.~\ref{fullsoln}
via differentiation:
\begin{equation}
    P(t|z)
    =-\frac{\partial  S(t|z)}{ \partial t} 
    =\sum_{n=1}^\infty \lambda_nc_n \psi_n(z) e^{-\lambda_n t}.
    \label{EQ24}
\end{equation}
% 
%where $\lambda_n=2\alpha_n k/\zeta$.,
The mean first-passage time (MFPT), $\mu(z)$, then follows from Eq.~\ref{EQ24}:
\begin{equation}
    \mu(z)
    =\int_0^\infty t P(t|z)\,dt
    =\sum_{n=1}^\infty \frac{c_n}{\lambda_n} \psi_n(z).
    \label{MFPTsoln}
\end{equation}

\section{Simulation}
\label{methods}

To test our theoretical results for the survival probability of a harmonically trapped particle diffusing outside a spherical, absorbing boundary, 
we performed simulations using the prescription given in Ref.~\cite{gillespie1996}.
Specifically,
the update formula for the $x$-coordinate of the bead at time $t+\Delta t$, namely
$x(t+\Delta t)$, in terms of its $x$-coordinate at time $t$, $x(t)$,
is given by
\begin{equation}
    x(t+\Delta t)=x(t)e^{-\frac{k}{\zeta}\Delta t}
    +\text{\bf{N}}(0,1)\sqrt{\frac{D \zeta}{k}(1-e^{-\frac{2 k}{\zeta}\Delta t})},
    \label{updateFormula}
\end{equation}
where $\text{\bf{N}}(0,1)$ denotes the standard Gaussian distribution.
Eq.~\ref{updateFormula} is valid for arbitrary $\Delta t>0$,
with identical expressions for the $y$- and $z$-coordinates.

To simulate the survival probability, $S(t|r_0)$,
we initialize
the particle's coordinates to be $(r_0,0,0)$
and the time to zero.
At each time step,
the particle's position is updated via Eq.~\ref{updateFormula}, applied 
independently for each of $x$, $y$, and $z$, and the resultant
separation between the particle and the origin is compared to the contact radius, $L$.
If the particle remains outside of the contact radius $L$, the elapsed time is incremented by $\Delta t$, and another update proceeds.
This process is repeated until the particle is inside the contact radius.
If the particle is inside the contact radius,
a capture event is considered to have occurred, and
the total elapsed time is recorded as one instance of the FPT.
The particle's position is then re-initialized to $(r_0,0,0)$, the elapse time is reset to zero, and the simulation is repeated 
until a total of $1.2\times10^6$ FPT measurements are collected.
% The whole process 
The parameters used in simulations are summarized in Table~\ref{Rouseparam}.
% lists the parameters used in the simulations.
% ???How many measurements of FPT? etc. etc.

\begin{center}
    \begin{table}
    \begin{tabular}{wl{5.0cm}wr{3.38cm}}
        \hline
        \hline
        Simulation parameter & Typical value \\
        \hline
        friction coefficient, $\zeta$ & $2.02$ nN$\cdot\mu$s/nm \\
        spring constant, $k$ & $(0.2525,1.01,2.02,4.04)$ fN/nm \\
        dimensionless trap stiffness, $\kappa$ & (0.003, 0.012, 0.024, 0.049) \\
        diffusivity, $D=\frac{k_BT}{\zeta}$ & 0.002 nm$^2/\mu$s\\
        temperature, T & 300 K\\
        relaxation rate, $\frac{k}{\zeta}=\tau^{-1}$ & (0.125, 0.5, 1, 2)~Hz\\
        initial separation, $r_0$ & 20, 50, 100, 200 nm\\
        contact distance, $L$ & 10 nm\\
        relative initial separation, $z=r_0/L$ &  2, 5, 10, 20 \\
        update time step, $\Delta t$ &  1 $\mu$s \\
        % simulation length & xxx \\
        \hline
        \hline
    \end{tabular}
    \caption{\label{Rouseparam} 
    Simulation parameters of a harmonically trapped particle.
    The friction coefficient and spring constant are the only control parameters used to update the particle's position, according to Eq.~\ref{updateFormula}.
    Values in parentheses indicate the specific parameters used to generate the simulation results shown in Figs.~\ref{fig:1}, \ref{fig:1.1}, and \ref{fig:2}.
    }
    \end{table}
\end{center}
% 

% To test the accuracy of our series solutions for the survival probability (Eq.~\ref{fullsoln}) and MFPT (Eq.~\ref{MFPTsoln}) with finite number of terms, 

\section{Results}
\label{Results}

\begin{figure}[htp]
    \includegraphics[width=0.45\textwidth]{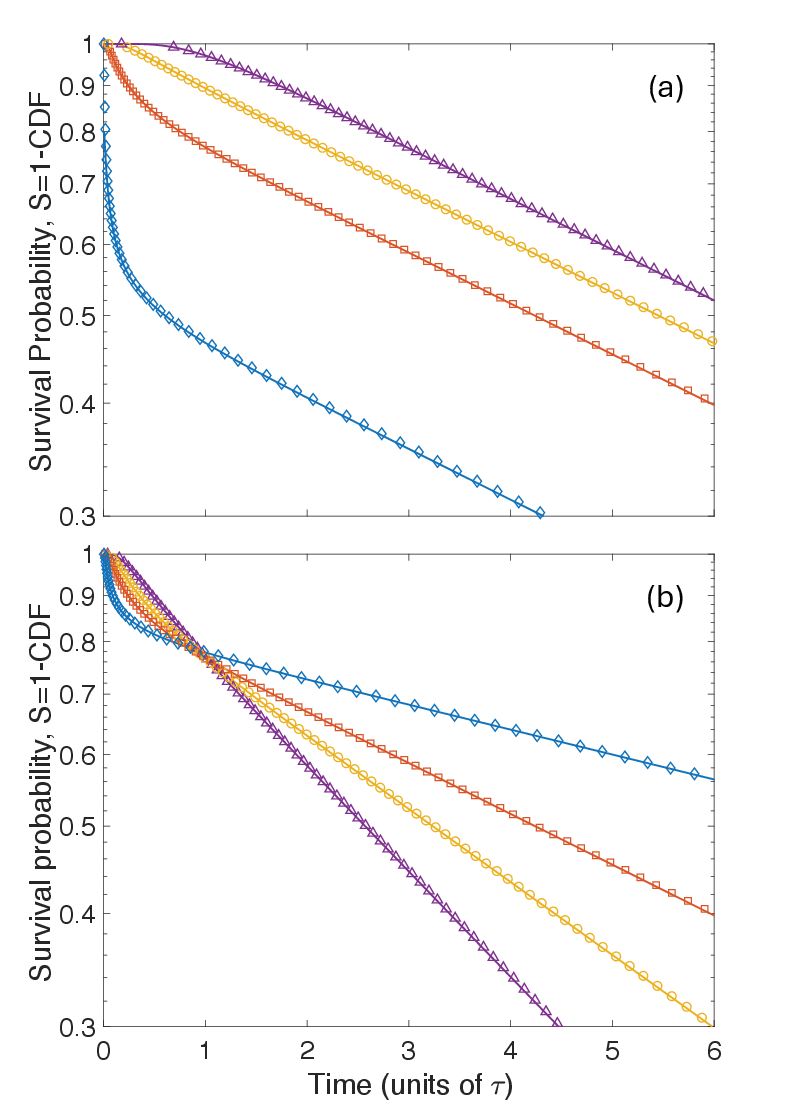}
    \caption{
    Survival probability as a function of time.
    (a) Different values of the dimensionless initial separation, $z$, with the
    dimensionless trap stiffness fixed at $\kappa=0.012$.
    From top to bottom, the ratio between initial separation and contact distance, $z$, is 
    20 (purple, triangle), 10 (yellow, circle), 5 (red, square), and 2 (blue, diamond).
    (b) Different values of $\kappa$ with $z$ fixed at 5.
    From top to bottom (as viewed from the right end of the plot),
    the value of $\kappa$ is
    0.003 (blue, diamond), 0.012 (red, square), 0.024 (yellow, circle), and 0.049 (purple, triangle).
    All lines represent the theoretical series solution  using the first twenty-five terms of Eq.~\ref{fullsoln},
    while the symbols depict the corresponding simulation results.
    In both panels, the time axes is in the units of the
    particle's relaxation time, $\tau=\frac{\zeta}{k}$.
    }
    \label{fig:1}
\end{figure}

Figure~\ref{fig:1}(a)  displays the survival probability, $S(t|z)$, on a logarithmic scale,
versus elapsed time, $t$, on a linear scale
for several values of the particle's initial radial coordinate ($z = 2, 5, 10, 20$, and $\kappa=0.012$).
The lines in Fig.~\ref{fig:1}(a) correspond to our theoretical results, given by Eq.~\ref{fullsoln}. The symbols correspond to the simulated survival probability,
calculated as the complement of the cumulative probability that the 
simulated FPT is less than a time, $t$.
Each simulated curve (presented as symbols) comprises a total of $1.2\times10^6$ measurements of the FPT
for each condition studied.
Evidently, there is excellent agreement between the theoretical results and the simulation results, giving confidence that our theoretical results are
indeed correct.
It is clear that for large values of $t$, all curves decrease linearly in this plot, corresponding to a simple exponential decay with the same time constant,
irrespective of the initial position.
At short times, however, the curves
for different initial coordinates differ significantly.
When the particle is initially far from the contact radius ($z=20$),
the survival probability remains close to unity for an apparent ``incubation'' time, before
eventually achieving the common long-time single-exponential decay.
By contrast, when the particle is initially close to the contact radius ($z=2$), the survival
probability falls precipitously  with increasing time before
its decrease slows to the common single-exponential decay.
% To examine the critical range of values of $z$ that bridges the two regimes described above,
% we plot the survival probability as a function of time for the initial radial coordinate of $z=6.4$ (dashed black line in Fig.~\ref{fig:1}), equal to the root mean squared length of the relaxed spring with dimensionless stiffness of $\kappa=0.012$.
For the intermediate case, the yellow line, corresponding to
an initial separation
of
$r_0=100$~nm ($z=10$), exhibits a nearly single exponential behavior across all times.
Interestingly,
an initial radial separation of $r_0=100$~nm, 
is close to
the particle's root mean squared displacement
from the origin,
$\sqrt{\langle r^2\rangle}=\sqrt{\frac{3k_BT}{k}} \approx111$~nm,
its
mean displacement
from the origin,
$\langle |r|\rangle=\sqrt{\frac{8}{\pi}}\sqrt{\frac{k_B T}{k}}\approx102$~nm,
and the mode of its displacement,
$r_{\text{mode}}=\sqrt{\frac{2k_B T}{k}}\approx91$~nm).
It seems natural,
therefore, 
to interpret the single-exponential behavior of the survival probability
at long times, 
which all initial separations
have in common, as reflecting the first-passage process from
the particle's equilibrium separation to the contact radius. 
This observation
suggest that we can envision the first-passage process from arbitrary initial conditions as follows.
For initial separations larger than the
equilibrium (mean) separation,
the observed ``incubation'' time prior to the
emergence of
the common single-exponential decay reflects the time required for the system to relax to its equilibrium state. 
Once the particle has
achieved its equilibrium separation,
subsequently,
it undergoes a single-exponential
first-passage process from the equilibrium separation to the capture distance.
On the other hand,
for initial separations less than the equilibrium
separation, the particle is either rapidly captured,
corresponding to the rapid initial decrease of the survival probability in this case, or
it temporarily escapes as far as the equilibrium separation, whereupon it
undergoes a single-exponential
first-passage process from the equilibrium separation to the capture distance.
% the incubation time and precipitous time...
% relaxation to the exponential behavior indicates the %relaxation o initially-far and initially-close regime
%For initial separations between $L$ and approximately $%\langle|R|\rangle$, 
%a fraction of the ensemble reaches the contact distance without undergoing stage (1), resulting in a precipitous, non-exponential drop in the survival probability at early times, as seen in the blue ($z=2$) and red ($z=5$) curves. 

% Therefore, we ascribe the change in regime to the starting radial coordinate relative to the root mean squared length of the relaxed spring.
% First-passage process can be viewed as a two-step process: 
Figure~\ref{fig:1}(b) shows the survival probability, $S(t|z)$, on a logarithmic scale,
versus elapsed time, $t$, on a linear scale
for an initial radial coordinate of $z=5$ and
several values of the dimensionless trap stiffness ($\kappa=0.003,0.012,0.024,0.049$).
Once again, we observe excellent agreement between the theoretical prediction and simulation results across all values of $\kappa$ shown. 
For large values of $t$, all curves decrease linearly with different slopes, corresponding to exponential decay with time constants that decrease as $\kappa$ increases. 
% It is interesting to examine the early time ($t\lesssim\tau$) behavior.
At early times, 
curves with smaller $\kappa$ exhibit a more rapid decline in survival probability compared to those with larger $\kappa$, 
which may seem counterintuitive.
This behavior arises because the relaxation time, $\tau$, increases as $\kappa$ decreases, so the same dimensionless time corresponds to a longer actual time for smaller $\kappa$.
% This explains the early time behavior: small $\kappa$ has precipitously decreasing survival probability compared to large $\kappa$, this is because $\tau$ is large for small $\kappa$. 

% 
\begin{figure}[htp]
    \includegraphics[width=0.45\textwidth]{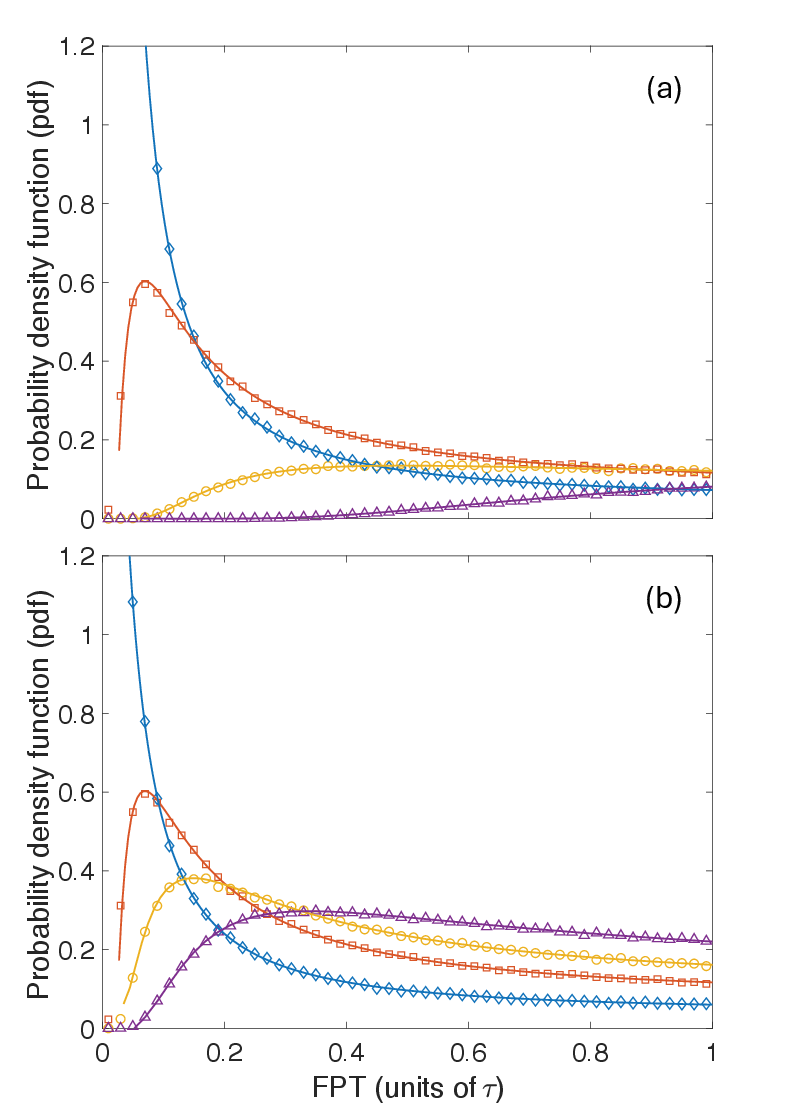}
    \caption{
    First-passage time distribution corresponding to the survival probability shown in Fig.~\ref{fig:1}, 
    using matching colors and symbols.
    This figure focuses on the early-time behavior ($\leq\tau$).
    The theoretical results (solid curves) are computed using the first fifty terms of Eq.~\ref{fullsoln}.
    Theoretical results are omitted for FPT $\lesssim$ 0.03~$\tau$
    due to large errors in the series solution at very early times, 
    where the behavior is dominated by the contribution of  higher-order terms.
    % (a) Different values of the dimensionless initial separations, $z$, with the
    % dimensionless trap stiffness fixed at $\kappa=?$.
    % From top to bottom, the ratio between initial separation and contact distance, $z$, is 
    % 20 (purple), 10 (yellow), 5 (red), and 2 (blue).
    % (b) Different values of $\kappa$ with $z$ fixed at 5.
    % Solid lines correspond to the sum of the first 25 terms of Eq.~\ref{fullsoln}.
    % The symbols correspond to the corresponding simulation results.
    % From top to bottom (as viewed from the right end of the plot),
    % the value of $\kappa$ is
    % 0.003 (blue), 0.012 (red), 0.024 (yellow), and 0.049 (purple).
    % In both cases, the time axes is in the units of the
    % particle's relaxation time,   $\tau=\frac{\zeta}{k}$.
    }
    \label{fig:1.1}
\end{figure}

Figure~\ref{fig:1.1} shows the FPT probability density functions, {\em i.e.} the FPT distributions,
$P(t|z)$, directly corresponding to the survival probabilities shown in Fig.~\ref{fig:1} with matching colors and symbols.
%Focusing on the early-time behavior ($\leq\tau$), 
%both panels
This figure also shows
% generally
excellent agreement between the theoretical and the simulation results
across a range of initial separations and potential stiffnesses. 
In fact, at early times (FPT $\leq\tau$), to achieve sufficient accuracy for the theoretical results shown,
we used the first fifty terms in the series solution in
this case.
For 
FPTs $\lesssim$ 0.03~$\tau$ even fifty terms was not enough to avoid large errors, and so
theoretical results for FPTs less than $0.03~\tau$
are not shown.
%due to the large error of the series solution for very %early times, 
%where the behavior is dominated by the contribution of %higher-order terms.
% Solutions for the survival probability and FPT distribution at very early times are dominated by the later terms in the series solution.

%
\begin{figure}[htp]
    \includegraphics[width=0.45\textwidth]{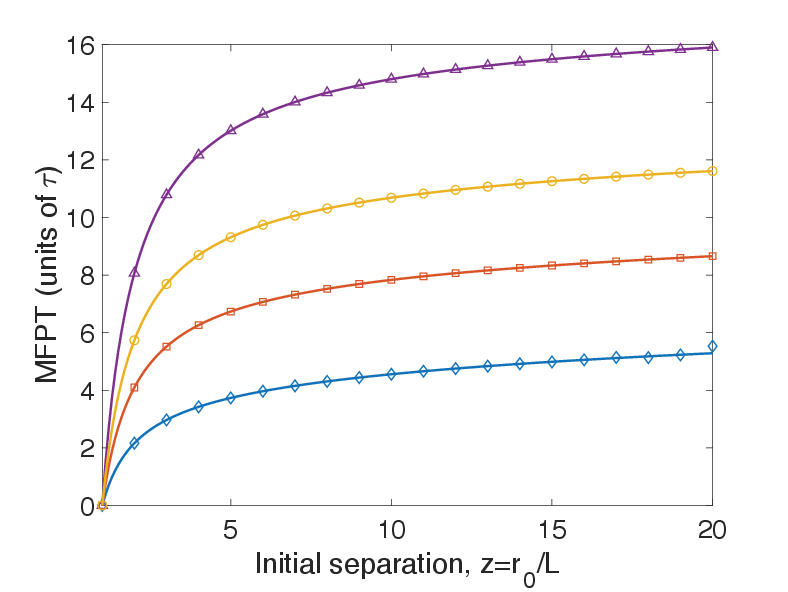}
    \caption{
    Mean first-passage time (MFPT), normalized by the particle's relaxation time, as a function of the initial separation, $z$, 
    shown for different values of $\kappa$.
    From top to bottom,
    the values of $\kappa$ are 0.003 (purple, triangle), 0.006 (yellow, circle), 0.012 (red, square), and 0.049 (blue, diamond).
    Solid curves represent the MFPT calculated using the integral representation by Grebenkov~\cite{grebenkov2014}, while symbols denote the series solution from Eq.~\ref{MFPTsoln}, truncated after the first twenty-five terms.
    }
    \label{fig:2}
\end{figure}

Figure~\ref{fig:2} shows the mean first-passage time (MFPT), $\frac{\mu}{\tau}$, normalized by the particle's relaxation time, as a function of the dimensionless initial radial coordinate, $z$, for several values of $\kappa$.
Results are presented for both
Grebenkov's integral solution (Eq.~82 in Ref.~\cite{grebenkov2014}, represented by solid curves in Fig.~\ref{fig:2} here) and our series solution (symbols) from Eq.~\ref{MFPTsoln}, using the first twenty-five terms.
The two solutions show excellent agreement across the range of $z$ and $\kappa$ values presented,
demonstrating that our truncated series solution with just twenty-five terms accurately reproduces the integral solution given by Grebenkov.
For small values of $z$ ($\lesssim5$), 
unsurprisingly, 
the normalized MFPT increases more rapidly with $z$ for smaller values of the trap stiffness, $\kappa$.
For large $z$, the normalized MFPT increases logarithmically with $z$, consistent with the behavior reported in Ref.~\cite{grebenkov2014}, and this long-time scaling is independent of $\kappa$.

\begin{figure}[htp]
    \includegraphics[width=0.45\textwidth]{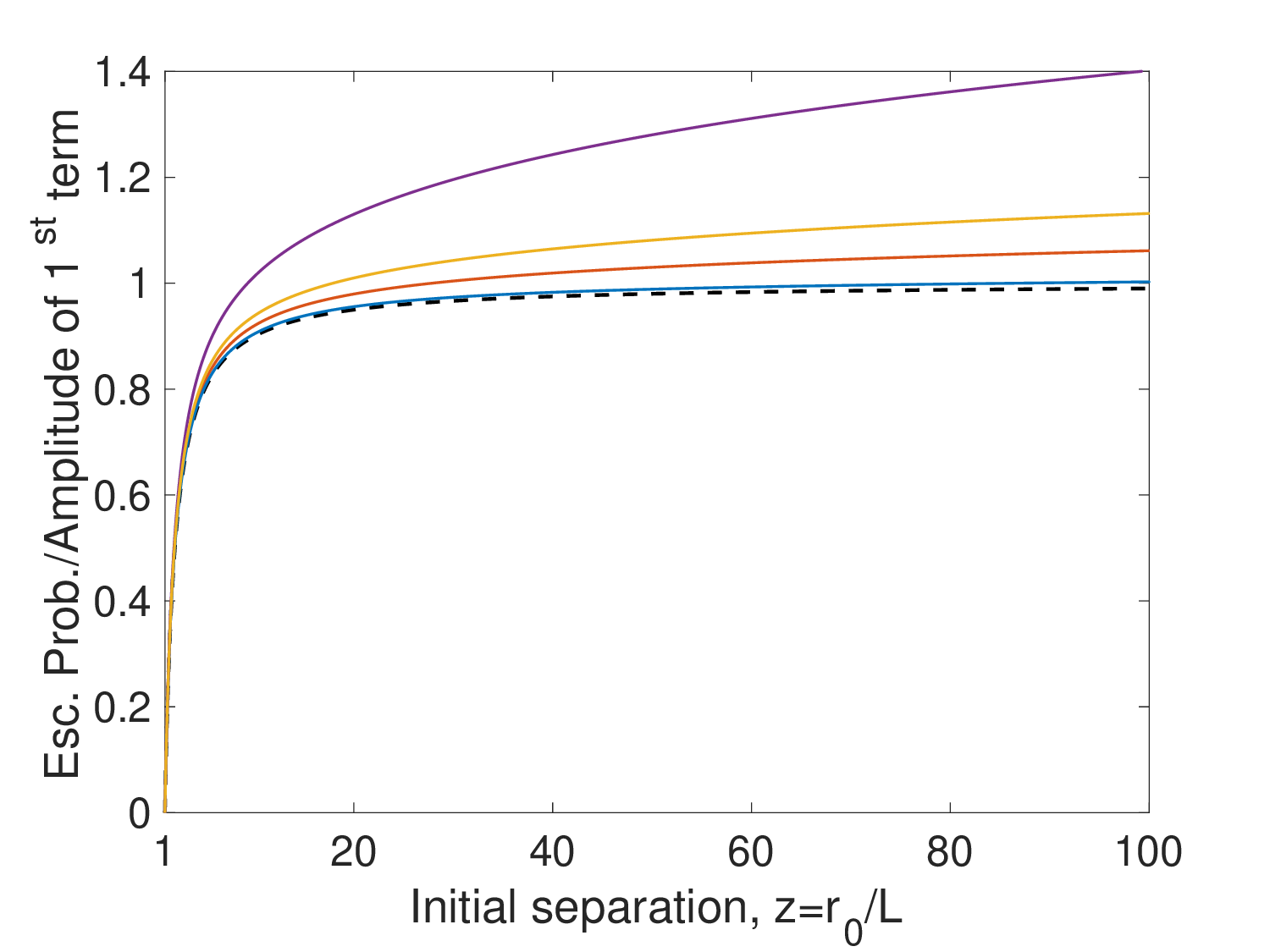}
    \caption{
    Theoretical escape probability, $P_\text{esc}(z)$, for the potential-free case (black dashed line), compared with the amplitudes of the first term, $c_1\psi_1(z)$, in the series solution for the survival probability (Eq.~\ref{fullsoln}), 
    shown for several small values of $\kappa$ approaching zero (solid lines). 
    From top to bottom, $\kappa$ for the solid lines are 0.012 (purple), 0.003 (yellow), 0.0012 (red), and 0.00012 (blue).
    % The black dashed line represents the theoretical escape probability for the potential-free diffusion-to-capture process.
    }
    \label{fig:3}
\end{figure}

The series solution for the survival probability given in Eq.~\ref{fullsoln}
is inapplicable when $\kappa=0$ (potential-free), as previously noted by Grebenkov~\cite{grebenkov2014}.
Mathematically, this is
because all of
the eigenvalues, $\lambda_n$, are zero for $\kappa=0$.
Physically, it is
because, in the absence of a
confining potential, the particle now has a finite probability to escape capture altogether.
The escape probability, $P_E$, can be readily calculated~\cite{berg1992}, 
yielding
%and can be derived by %solving the diffusion %equation with the %appropriate boundary %conditions.
%A compact expression for %this solution can be found, %for example, in ~
% with the result:
% 
\begin{equation}
    P_E(z)=1-\frac{1}{z}.
\end{equation}
% 
% which can be obtained by solving the Laplace equation of $P_\text{esc}$ with boundary conditions $P_\text{esc}(z=1)=0$ and $P_\text{esc}(z=\infty)=1$.
Nevertheless,
Equation~\ref{fullsoln} is valid for infinitesimal, positive $\kappa$,
in which case
evaluation of
Eq.~\ref{fullsoln} reveals that  the long-time limit of the survival probability versus time,
namely
$c_1 \psi_1 e^{\frac{2 k}{\zeta} \alpha_1 t}$, shows an ever slower decay as $k\rightarrow 0^+$, approaching a constant equal to $c_1 \psi_1$.
%Interestingly,
%reflecting the amplitude of the escape probability, $P_\text{esc}(z)$. 
%Since the term corresponding to the smallest zero of the Tricomi function (the %first term in Eq.~\ref{fullsoln}, by our definition) governs the long-time behavior,
%the amplitude of this first-order term, $c_1\psi_1(z)$, should approach to %$P_\text{esc}(z)$ in the limit of $\kappa\rightarrow 0$.
%
% On the other hand, when $\kappa=0$, 
% the theoretical escape probability,
% $P_\text{esc}$, admits a simple expression as a function of $z$:
% % 
% \begin{equation}
%     P_\text{esc}(z)=1-\frac{1}{z},
% \end{equation}
% % 
% which is obtained by solving the Laplace equation of $P_\text{esc}$ with boundary conditions $P_\text{esc}(z=1)=0$ and $P_\text{esc}(z=\infty)=1$.
Figure~\ref{fig:3} compares   $P_E(z)$ (dashed line) to $c_1\psi_1(z)$ (solid lines) as a function of $z = r_0/L$
for several (small) values of $\kappa$.
Interestingly, we see that $c_1 \psi_1$
progressively
approaches $P_E$ as $\kappa\rightarrow 0^+$ across values of $z$ studied.

\section{Conclusions}
\label{Conclusions}

By directly solving the differential equation for the survival probability, 
we have derived analytic series-form solutions for the survival probability, first-passage time (FPT) distribution, and mean first-passage time (MFPT) of a harmonically trapped particle diffusing outside an absorbing spherical boundary.
Our series solution corrects earlier results for the survival probability reported by Grebenkov~\cite{grebenkov2014}.
Through comparison with simulations of a harmonically trapped particle, we have shown that 
the series solution---using only the first twenty-five terms---accurately reproduces both the survival probability and FPT distribution across a broad range of initial separations, $z$, and trap stiffnesses, $\kappa$, for times greater than approximately 0.2~$\tau$.
For earlier times ($t\lesssim 0.2~\tau$), 
additional terms are required for accurate predictions.
We have also demonstrated that the truncated series solution for the MFPT, with just twenty-five terms, closely matches Grebenkov's integral solution across a range of $z$ and $\kappa$.
By examining our series solution of the survival probability in the limit of $\kappa$ approaching to zero,
we have found that 
the amplitude of the first term in the series solution approaches to the
theoretical escape probability for the potential-free diffusion-to-capture process.

% In spite of the simplicity in its concept, 
% the FPT analysis of a harmonically trapped particle provides the theoretical framework for key processes in physics, chemistry, and biology. 
% More specifically, 
% dynamics of end-to-end vector of a Rouse polymer in the time scale that is much larger than the single-bead relaxation time.

A Mathematica notebook for analytically computing the survival probability as a function of time and the MFPT as a function of $z$, based on the series-form solution, is available on GitHub~\cite{FPT_bead_code}. 
In addition, the MATLAB code used to perform FPT simulations of a harmonically trapped particle is also provided on GitHub~\cite{FPT_bead_code}.

\begin{acknowledgments}
This research was supported by the NSF Physics of Living Systems via Award 2412859.
\end{acknowledgments}
%

% \appendix
%
% \section{aaa}

\vspace{4mm}

\bibliography{reference_FPT}
\end{document}